\documentclass[aps,prc,superscriptaddress,nofootinbib,showpacs,floatfix,singlecolumn]{revtex4}
\usepackage{url}
\usepackage{cancel}
\usepackage[colorlinks,linkcolor=blue,citecolor=blue,filecolor=black,urlcolor=blue]{hyperref}
\usepackage{epsfig,graphics}
\usepackage{graphicx}% Include figure files
\usepackage{dcolumn}% Align table columns on decimal point
\usepackage{bm}% bold math
\usepackage[usenames]{color}
\usepackage{amssymb}
\usepackage{amsmath}
\usepackage{multirow}
\usepackage{float}
\usepackage{harpoon}
\usepackage{MnSymbol}
\usepackage{appendix}
\usepackage{color}
\usepackage{hyperref}
\usepackage{cleveref}

\newcommand{\Nch} {N_{\mathrm{ch}}}
\newcommand{\sqrtsnn}{\mbox{$\sqrt{s_{\mathrm{NN}}}$}}
\newcommand{\pT} {p_{\mathrm{T}}}
\newcommand{\lr}[1]{\left\langle #1\right\rangle}

\newcommand{\Dphi}{\mbox{$\Delta \phi$}}
\newcommand{\Deta}{\mbox{$\Delta \eta$}}
\newcommand{\nch}{dN_{\mathrm{ch}}/d\eta}

\newcommand{\rhon}[1] {\rho(v_{ #1 }^2,[\pT])}
\newcommand{\cov}[1] {\mathrm{cov}(v_{ #1 }^2,[\pT])}
\newcommand{\var}[1] {\mathrm{var}(v_{ #1 }^2)}
\newcommand{\varp} {\mathrm{var}([\pT])}
\newcommand{\bq} {\textbf{q}}
\newcommand{\bo} {\textbf{o}}

\newcommand{\HIJING}{{\tt Hijing}}

\newcommand{\PYTHIA}{{\tt Pythia8}}

\newcommand{\ptk}[1] {p_{\mathrm{T},#1}}

\usepackage{CJK}
\hfuzz=\maxdimen
\begin{document}
\title{Non-flow effects in correlation between harmonic flow and transverse momentum in nuclear collisions}
\newcommand{\sbu}{Department of Chemistry, Stony Brook University, Stony Brook, NY 11794, USA}
\newcommand{\bnl}{Physics Department, Brookhaven National Laboratory, Upton, NY 11976, USA}
  \author{Chunjian Zhang}\affiliation{\sbu}
  \author{Arabinda Behera}\affiliation{\sbu}
\author{Somadutta Bhatta}\affiliation{\sbu}
\author{Jiangyong Jia}\email[Correspond to\ ]{jiangyong.jia@stonybrook.edu}\affiliation{\sbu}\affiliation{\bnl}
\date{\today}
\begin{abstract}
A large anti-correlation signal between elliptic flow $v_2$ and average transverse momentum $[p_{\mathrm{T}}]$ was recently measured in small collision systems, consistent with a final-state hydrodynamic response to the initial geometry. This negative $v_2$--$[p_{\mathrm{T}}]$ correlation was predicted to change to positive correlation for events with very small charged particle multiplicity $N_{\mathrm{ch}}$ due to initial-state momentum anisotropies of the  gluon saturation effects. However, the role of non-flow correlations is expected to be important in these systems, which is not yet studied. We estimate the non-flow effects in $pp$, $p$Pb and peripheral PbPb collisions using {\tt Pythia} and {\tt Hijing} models, and compare them with the experimental data. We show that the non-flow effects are largely suppressed using the rapidity-separated subevent cumulant method (details of the cumulant framework are also provided). The magnitude of the residual non-flow is much less than the experimental observation in the higher $N_{\mathrm{ch}}$ region, supporting the final-state response interpretation. In the very low $N_{\mathrm{ch}}$ region, however, the sign and magnitude of the residual non-flow depend on the model details. Therefore, it is unclear at this moment whether the sign change of $v_2$--$[p_{\mathrm{T}}]$ can serve as evidence for initial state momentum anisotropies predicted by the gluon saturation.
\end{abstract}

%	25.75.Gz, Particle correlations and fluctuations
%	25.75.Ld,	Collective flow, relativistic collisions.
%      25.75.-q,	Relativistic heavy-ion collisions                
\maketitle

\section{Introduction}
\label{introduction}
In high-energy hadronic collisions, particle correlations are an important tool to study the multi-parton dynamics of QCD in the strongly coupled non-perturbative regime~\cite{Shuryak:2014zxa}. Measurements of azimuthal correlations in small collision systems, such as $pp$ and $p$+A collisions~\cite{CMS:2012qk,Abelev:2012ola,Aad:2012gla,Aad:2014lta,Khachatryan:2015waa}, have revealed a strong harmonic modulation of particle densities d$N/{\textrm d}\phi\propto 1+2\sum_{n=1}^{\infty}v_{n}\cos n(\phi-\Phi_{n})$. Measurement of $v_n$ and their event-by-event fluctuations have been performed as a function of charged particle multiplicity $\Nch$ in $pp$ and $p$+A collisions. It is found that the azimuthal correlations involve all particles over a wide pseudorapidity range. A key question is whether this multi-particle collectivity reflects initial momentum correlation from gluon saturation effects (ISM)~\cite{Dusling:2013qoz}, or a final-state hydrodynamic response to the initial transverse collision geometry (FSM)~\cite{Bozek:2013uha}.

Recently, the correlation between $v_n$ and $[\pT]$, the average transverse momentum of particles in each event, was proposed to be a sensitive observable to distinguish between the initial-state and final-state effects~\cite{Giacalone:2020byk}. The lowest order of such correlation is characterized by the covariance $\cov{n}\equiv \lr{v_n^2[\pT]} - \lr{v_n^2}\lr{[\pT]}$~\cite{Bozek:2016yoj} with the average carried over events, which have been measured at the LHC~\cite{Aad:2019fgl,ATLAS:2021kty}. In the final-state dominated scenario, the flow harmonics are diven by the initial spatial eccentricity $\varepsilon_n$, $v_n\propto\varepsilon_n$, while the $[\pT]$ is related to the transverse size of the overlap region: events with similar total energy but smaller transverse size in the initial state are expected to have a stronger radial expansion and therefore larger $[\pT]$~\cite{Bozek:2012fw}.  Hydrodynamic model calculations with negligible initial transverse momentum predict a positive $\cov{n}$ at large $\Nch$ which changes to negative $\cov{n}$ towards small $\Nch$ region\cite{Bozek:2020drh,Giacalone:2020dln,Schenke:2020uqq}, whereas at small enough Nch, initial momentum anisotropy can, in fact, dominate. In a gluon saturation picture, these correlations are expected to give a positive contribution to $\cov{n}$~\cite{Giacalone:2020byk}. Therefore, the $\Nch$ dependence of $\cov{2}$, after considering both initial and final-state effects, is predicted to exhibit a double sign change as a function $\Nch$. The experimental observation of such sign change was further argued to provide a strong evidence for the gluon saturation physics~\cite{Giacalone:2020byk}.

On the other hand,  momentum correlations could also arise from ``non-flow'' effects from resonance-decays, jets and dijets~\cite{Jia:2013tja}. Such non-flow correlations usually involve a few particles from one or two localized pseudorapidity regions, in contrast to the initial momentum correlation from gluon saturation, which spans continuously over a large rapidity range similar to hydrodynamic flow. The non-flow effects are often suppressed by correlating particles from two or more subevents separated in pseudorapidity. This so-called subevent cumulant method~\cite{Jia:2017hbm} has been validated for several multi-particle correlators involving flow harmonics of same or different orders\cite{Jia:2017hbm,Huo:2017nms,Zhang:2018lls}, such as four-particle cumulants $c_{n}\{4\}=\left\langle v_{n}^{4}\right\rangle-2\left\langle v_{n}^{2}\right\rangle^{2}$, four-particle symmetric cumulants $\left\langle v_{n}^{2} v_{m}^{2}\right\rangle-\left\langle v_{n}^{2}\right\rangle\left\langle v_{m}^{2}\right\rangle$ and three particle asymmetric cumulants $\left\langle v_{n} v_{m} v_{n+m} \cos \left(n \Phi_{n}+m \Phi_{m}-(n+m) \Phi_{n+m}\right)\right\rangle$. It is found that results from the standard cumulant method are contaminated by non-flow correlations in $pp$, $p$A and peripheral AA collisions, while they are largely suppressed in the subevent method that requires three or more subevents~\cite{Aaboud:2017blb,Aaboud:2017blb,Aaboud:2019sma}. Since covariance $\cov{n}$ is a three-particle correlator, it can be measured with two subevent or three-subevent methods, which suppress the non-flow while keeping the genuine long range multi-particle correlations associated with ISM and FSM.
%hydrodynamic response. 

In this paper, we study the influence of the non-flow correlations to covariance $\cov{n}$ in $pp$, $p$Pb and PbPb collisions using $\PYTHIA$~\cite{Sjostrand:2007gs} A2 tune and $\HIJING$ v1.37~\cite{Gyulassy:1994ew} models in the standard and subevent methods. We find that the non-flow correlations give a positive contribution to $\cov{2}$, which are strongly suppressed in the three-subevent method, but not completely eliminated. The sign and magnitude of the residual non-flow are model dependent. Therefore, the mere observation of change of $\cov{2}$ from negative to positive towards low $\Nch$ in the experimental results may not serve as evidence for the presence of gluon saturation.

\section{Methodology and model setup}\label{sec:2}
The covariance $\cov{n}$ is a three-particle correlator, which is obtained by averaging over unique triplets in each event, and then over all events in an event class~\cite{Bozek:2016yoj,Bozek:2020drh}:
\begin{align}\label{eq:1}
\cov{n} &= \lr{\frac{\sum_{i,j,k, i\neq j\neq k} w_iw_jw_k e^{in(\phi_i-\phi_j)}(\ptk{k}-\lr{ [\pT]})} {\sum_{i,j,k, i\neq j\neq k} w_iw_jw_k}}
\end{align}
where the indices $i$, $j$ and $k$ loop over distinct charged particles to account for all unique triplets, the particle weight $w_i$ is constructed to correct for detector effects, and the $\lr{}$ denotes average over events. In order to reduce short-range ``non-flow'' correlations, pseudorapidity gaps are often explicitly required between the particles in each triplet. This analysis uses the so-called standard, two-subevent and three-subevent methods~\cite{Jia:2017hbm} to explore the influence of non-flow correlations as detailed below.

The choices of $\eta$ ranges for the subevents are identical to those used by the ATLAS experiment~\cite{Aad:2019fgl,ATLAS:2021kty}. In the standard method, all charged particles within $|\eta|<2.5$ are used. In the two-subevent method, triplets are constructed by combining particles from two subevents labeled as $a$ and $c$ with a $\Deta$ gap in between to reduce non-flow effects: $-2.5<\eta_a<-0.75\;,\; 0.75<\eta_c<2.5$. The two particles contributing to the flow vector are chosen as one particle each from $a$ and $c$, while the third particle providing the $\pT$ weight is taken from either $a$ or $c$. In the three-subevent method, three non-overlapping subevents $a$, $b$ and $c$ are chosen: $-2.5<\eta_a<-0.75\;,|\eta_b|<0.5\;,\; 0.75<\eta_c<2.5$. The particles contributing to flow are chosen from subevents $a$ and $c$ while the third particle is taken from subevent $b$. 

A direct calculation of the nested-loop in Eq.~\eqref{eq:1} is computationally expensive. Instead, it can be expanded algebraically within the multi-particle cumulant framework~\cite{Bilandzic:2010jr,Jia:2017hbm} into a polynomial function of vectors and scalars:
\begin{align}\label{eq:2}
&\bq_{n;k}=\frac{\sum_i w_i^k e^{in\phi_i}}{\sum_i w_i^k}\;,\bo_{n;k}=\frac{\sum_i w_i^k e^{in\phi_i}(p_{\mathrm{T},i}-\lr{[\pT]})}{\sum_i w_i^k}\;,\;p_{m;k}=\frac{\sum_i w_i^k(p_{\mathrm{T},i}-\lr{[\pT]})^m}{\sum_i w_i^k}\;,\tau_k =\sum_i w_i^{k+1}/(\sum_i w_i)^{k+1}
\end{align}
where the sum runs over particles in a given event or subevent and ``$k$'' and ``$m$'' are natural integer powers. It is straightforward to show that expansion of Eq.~\eqref{eq:1} in the three methods gives:
\begin{align}\label{eq:3a}
\cov{n}_{\mathrm{std}}  &= \lr{\frac{(|q_{n;1}|^2-\tau_{1})p_{1;1}-2\tau_{1}\Re(\bo_{n;2}\bq_{n;1}^*)+2\tau_{2}p_{1;3}}{1-3\tau_{1}+2\tau_{2}}} \\\label{eq:3b}
\cov{n}_{\mathrm{2sub}} &=\lr{\frac{\Re[(\bq_{n;1}p_{1;1}-\tau_{1}\bo_{n;2})_a (\bq_{n;1}^*)_c+(\bq_{n;1}p_{1;1}-\tau_{1}\bo_{n;2})_c (\bq_{n;1}^*)_a] }{1-(\tau_{1})_a +1-(\tau_{1})_c}}\\\label{eq:3c}
\cov{n}_{\mathrm{3sub}} &=\lr{\Re[(\bq_{n;1})_a(\bq_{n;1}^*)_c] (p_{1;1})_b}
\end{align} 
where the $\Re$ denotes the real component of the complex number.

Experimentally, the $v_n$ -$[\pT]$ correlation is aften presented in normalized form known as Pearson's correlation coefficient~\cite{Bozek:2016yoj},
\begin{align} \label{eq:4}
\rhon{n} =\frac{\cov{n}}{\sqrt{\var{n}}\sqrt{\varp}}\;,
\end{align}
where the $\varp$ and $\var{n}$ are variances of $\pT$ fluctuations and $v_n^2$ fluctuations, respectively. The $\varp$ is obtained using all the pairs in the full event $|\eta|<2.5$,
\begin{align}\label{eq:5}
 \varp=\lr{\frac{\sum_{i,j, i\neq j} w_iw_j\ptk{i}-\lr{[\pT]})(\ptk{j}-\lr{[\pT]})} { \sum_{i,j,i\neq j} w_iw_j}}=\lr{\frac{p_{1;1}^2-p_{2;2}}{1-\tau_{1}}}
\end{align}
The dynamical variance $\var{n}$ are calculated in terms of two-particle cumulant $c_n\{2\}$ and four particle cumulants $c_n\{4\}$ following Ref.~\cite{ATLAS:2021kty}:
\begin{align}\label{eq:6}
\var{n} \equiv \lr{v_n^4}-\lr{v_n^2}^2 =  c_n\{4\}_{\mathrm{std}} + c_n\{2\}^2_{\mathrm{2sub}}\;.
\end{align}
The $c_n\{4\}$, being a four-particle correlator, is known to be relatively insensitive to non-flow correlations but usually has poor statistical precision. Therefore it is obtained from the standard cumulant method using the full event. On the other hand, the two particle cumulants $c_n\{2\}$ is more susceptible to non-flow correlations and therefore is calculated from the two-subevent method with the $\eta$ choices discussed above. This definition is mostly free of non-flow in large collision systems. But in small systems, this definition could still be biased by non-flow effects as we discussed in Appendix~\ref{sec:app1}.

To evaluate the influence of non-flow correlations to $\cov{n}$ and $\rhon{n}$, the $\PYTHIA$ A2 tune~~\cite{Sjostrand:2007gs} and $\HIJING$ v1.37~\cite{Gyulassy:1994ew} models are used to generate $pp$ events at $\sqrt{s} = 13$ GeV, $p$Pb and peripheral PbPb events at $\sqrtsnn = 5.02$ TeV, respectively. These models contain significant non-flow correlations from jets, dijets, and resonance decays and can be used to quantify the efficacy of non-flow suppression in these methods.  In these simulations, the particle weight are set to be unity, $w_i=1$ and events are classified by $\Nch$, the number of charged particles in $|\eta|<2.5$ with $\pT>0.1$ GeV. The $\cov{n}$ are calculated in three $\pT$ ranges using the standard and subevent methods: $0.2<\pT<2$ GeV,  $0.5<\pT<2$ GeV, and $0.5<\pT<5$ GeV.  They are presented as a function of charged particle density at mid-rapidity $d\Nch/d\eta$, which is assumed to be 1/5 of $\Nch$, $\Nch\approx 5 d\Nch/d\eta$. 

\section{results} \label{results}
\begin{figure}[h!]
\begin{center}
\includegraphics[width=1.0\linewidth]{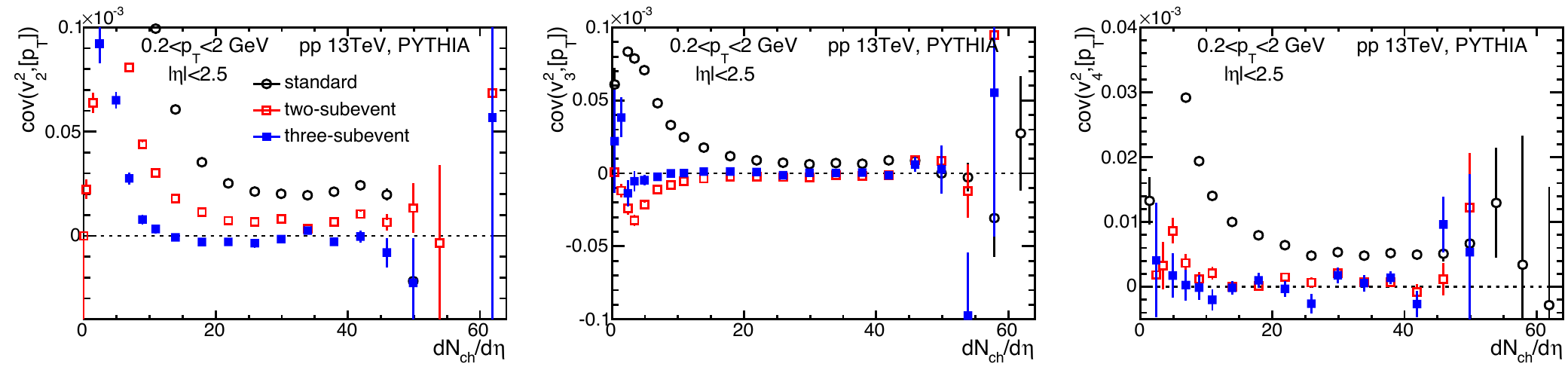}
\end{center}
\caption{\label{fig:1} The $\cov{n}$ as a function of $\nch$ for $n=2$ (left),3 (middle),4 (right) compared between the standard, two- and three-subevent methods for charged particles in $0.2 < \pT < 2$ GeV obtained from 13 TeV $pp$ $\PYTHIA$.}
\end{figure}

Figure~\ref{fig:1} compares the results of $\cov{n}$ from the standard and subevent methods in $pp$ collisions from $\PYTHIA$ model. The values from the standard method are positive for all harmonics. This is because the correlations are dominated by the jet fragmentations, which produce clusters of particles with larger $\pT$ and enhanced azimuthal correlations at $\Dphi\sim0$, and therefore tend to simultaneously increase the $v_n^2$ and $[\pT]$. The values from the two-subevent method are positive for even harmonics and negative for odd harmonics, consistent with the dominance of correlations from away-side jet fragments: the away-side correlations are expected to give a more negative $v_3^2$ and larger $[\pT]$, and therefore a negative value of $\cov{3}$. For the three-subevent method, the values of $\cov{2}$ are positive at $\nch\lesssim10$ and are slightly negative for $\nch>10$. The magnitudes of $\cov{n}$ are largest for the standard method, and smallest for the three-subevent method. Similar ordering among the three methods are observed in all three collision systems and all $\pT$ selections, and the magnitudes of signal from three-subevent method are always the smallest, suggesting that this method is least affected by non-flow. For the remaining discussion, we focus on discussing results from the three-subevent method.

\begin{figure}[h!]
\begin{center}
\includegraphics[width=1.0\linewidth]{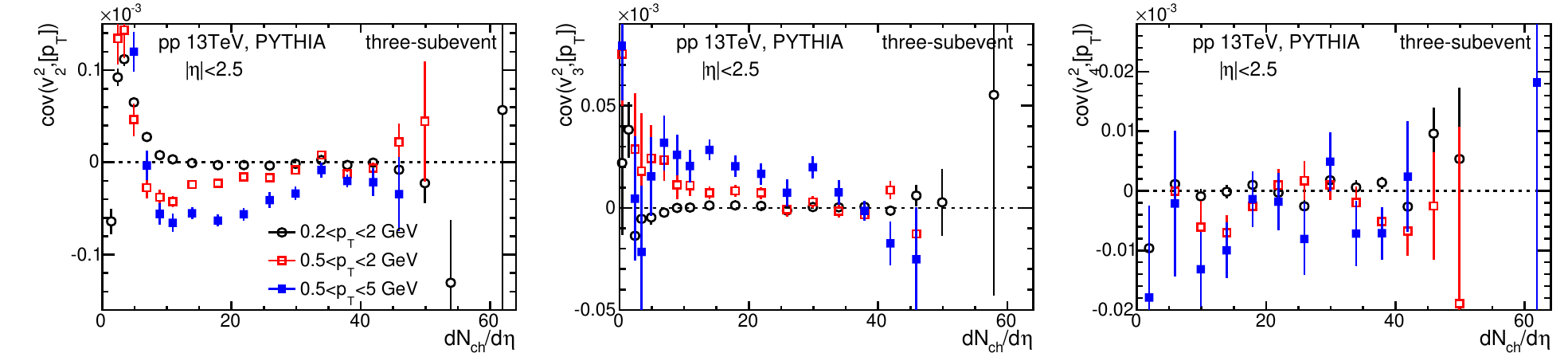}
\end{center}
\caption{\label{fig:2} The $\cov{n}$ as a function of $\nch$ from the three-subevent method for $n=2$ (left), 3 (middle),4 (right) in three $\pT$ ranges in 13 TeV $pp$ collisions.}
\end{figure}

Figure~\ref{fig:2} compares the results of $\cov{n}$ from three $\pT$ ranges. The overall magnitudes of $\cov{n}$  are larger in the higher $\pT$ range, reflecting a larger non-flow correlation at higher $\pT$. The values of $\cov{2}$ exhibit qualitatively a similar sign change behavior at $\nch\sim5-10$ for all $\pT$ ranges. The values of $\cov{3}$ are mostly positive, and the values of $\cov{4}$ seem to be systematically below zero.

\begin{figure}[h!]
\begin{center}
\includegraphics[width=1.0\linewidth]{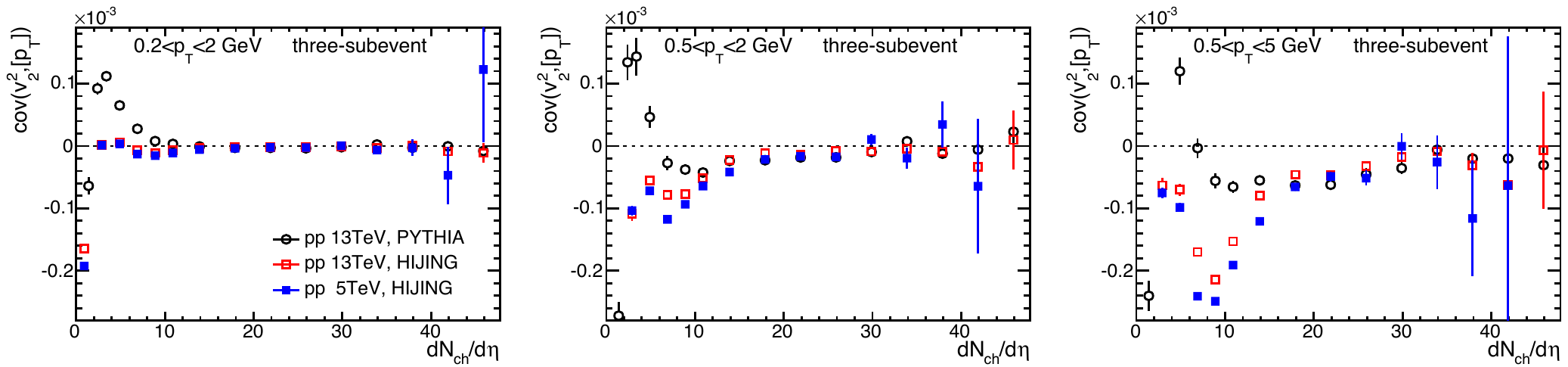}
\end{center}
\caption{\label{fig:3} The $\cov{2}$ as a function of $\nch$ from the three-subevent method compared between three pp collision systems for $0.2 < \pT < 2$ GeV (left), $0.5 < \pT < 2$ GeV (middle), and $0.5 < \pT < 5$ GeV (right).}
\end{figure}

To further investigate the origin of the sign change of $\cov{2}$ in the low $\nch$ region, Figure~\ref{fig:3} compares the $pp$ results from $\PYTHIA$ with those obtained from the $\HIJING$ model. The results are in good quantitative agreement for $\nch>20$. In the $\nch<30$ range and towards lower $\nch$, the $\HIJING$ results show a stronger decrease compared to the $\PYTHIA$ results. The $\HIJING$ results start to increase at $\nch<10$ similar to $\PYTHIA$, but except for the lowest $\pT$ range of $0.2<\pT<2$ GeV, the increase is not enough for the $\cov{2}$ to change sign. The results from $pp$ collisions at $\sqrt{s}=5$ TeV are also shown in Figure~\ref{fig:3}. The values are more negative than those for the $\sqrt{s}=13$ TeV results for $\nch<20$, suggesting that residual non-flow is larger at lower $\sqrt{s}$ at the same $\nch$.
 
\begin{figure}[h!]
\begin{center}
\includegraphics[width=1.0\linewidth]{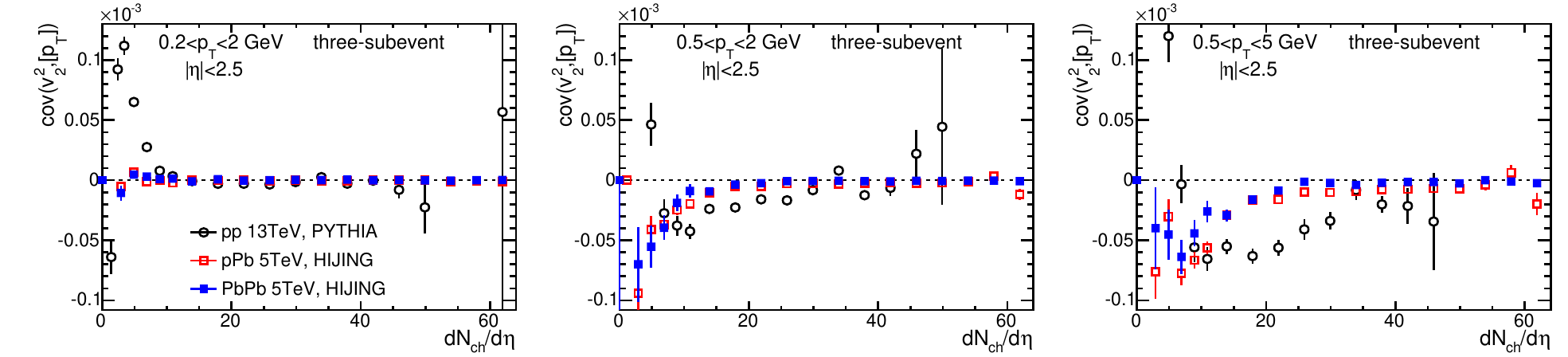}
\end{center}
\caption{\label{fig:4} The $\cov{2}$ as a function of $\nch$ from the three-subevent method compared between three collision systems for $0.2 < \pT < 2$ GeV (left), $0.5 < \pT < 2$ GeV (middle), and $0.5 < \pT < 5$ GeV (right).}
\end{figure}

Figure~\ref{fig:4} compares the results of $\cov{2}$ between $pp$, $p$Pb and PbPb collisions, separately in three $\pT$ ranges. The $p$Pb and PbPb values are negative at low $\nch$ region, whose magnitudes increase with $\pT$. This is different from the $pp$ results, which are positive at $\nch\lesssim8$ region. In the $\nch>10$ region, the $pp$ values are negative and lower than those for the $p$Pb and PbPb collisions. The values for $p$Pb collisions are close to but consistently lower than those in PbPb collisions, suggesting a slightly larger residual non-flow in $p$Pb collisions.

In order to estimate the non-flow effects on the $\rhon{n}$, we need to choose an appropriate normalization in Eq.~\eqref{eq:4}. The $\var{n}$ directly obtained from these models should not be used, because they only contain non-flow. Instead, we estimate $\var{n}$ from the previous published measurements of $v_n\{2\}$ and $v_n\{4\}$ in these three collision systems~\cite{Aaboud:2017blb,Aaboud:2018syf,Khachatryan:2015waa} as:
\begin{align}\label{eq:7}
\var{n} = \langle v_n^4\rangle -\langle v_n^2\rangle^2 =v_{n,\mathrm{tmp}}\{2\}^{4}\left(1-\left[\frac{v_n\{4\}}{v_n\{2\}}\right]^4\right)
\end{align}
The $v_{n,\mathrm{tmp}}\{2\}$ were measured using the two-particle correlation and improved template method from Ref.~\cite{Aaboud:2018syf} that explicitly subtracts the non-flow correlations. The $\pT$ dependence of the $v_{n,\mathrm{tmp}}\{2\}$ are taken from Ref.~\cite{Aaboud:2018syf}. The values of $v_2\{4\}/v_2\{2\}$ are taken from Ref.~\cite{Aaboud:2017blb} for $pp$ and $p$Pb and from Ref.~\cite{Khachatryan:2015waa} for PbPb, which are found to be in the range of 0.71--0.74 as a function of $\nch$, and they are assumed to be independent of $\pT$. The $v_2\{4\}$ term leads to a 28\% reduction to $\var{2}$. For third-order harmonics, the values of $v_3\{4\}/v_3\{2\}$ have been found to be very small Ref.~\cite{Aaboud:2019sma} and therefore is neglected in this study, i.e. we assume $\var{3}=v_{3,\mathrm{tmp}}\{2\}^{4}$. Examples of the $\nch$ dependence of $\var{2}$ and $\var{3}$ are given in Figure~\ref{fig:app1} of Appendix~\ref{sec:app1}.
\begin{figure}[h!]
\begin{center}
\includegraphics[width=0.9\linewidth]{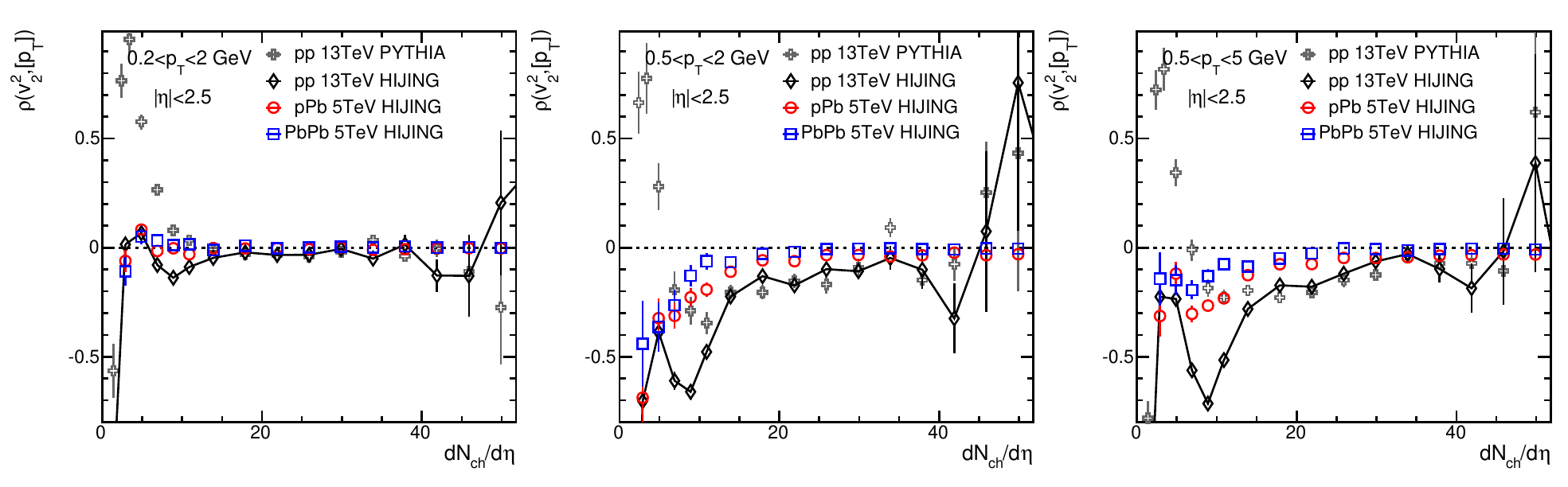}
\includegraphics[width=0.9\linewidth]{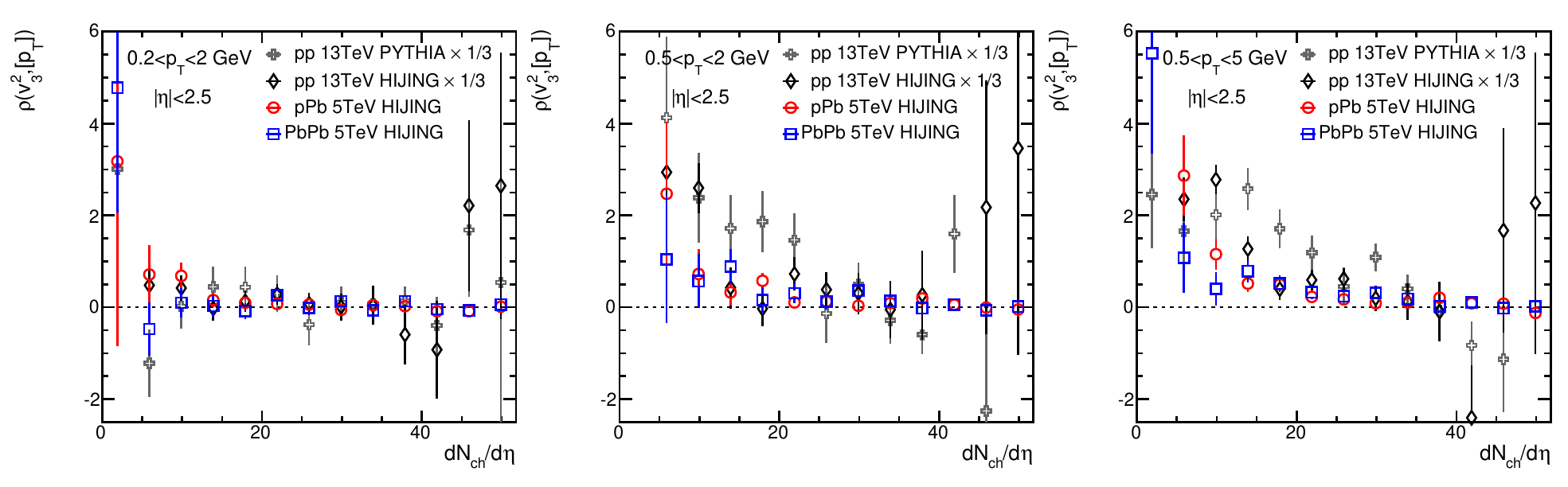}
\end{center}
\caption{\label{fig:5} The $\rhon{2}$ (top) and $\rhon{3}$ (bottom) estimated via as a function of $\nch$ from the three-subevent method compared between three collision systems for $0.2 < \pT < 2$ GeV (left), $0.5 < \pT < 2$ GeV (middle), and $0.5 < \pT < 5$ GeV (right). Note that the $\rhon{3}$ from $pp$ collisions have been scaled down by a factor of 3.}
\end{figure}

The results of $\rhon{2}$ and $\rhon{3}$ are shown in Figure~\ref{fig:5} for the three collision systems. They provide an estimate of the expected non-flow contributions to the experimentally measured $\rhon{n}$. In the $0.2<\pT<2$ GeV and $\nch>12$ region, the values of $|\rhon{2}|$ are $<0.02$ in $p$Pb and PbPb collisions and are $<0.05$ in the $pp$ collisions. An experimental observation of a signal much larger than these values could be a clear indication of non-trivial initial- and final-state correlations unrelated to non-flow. In the higher $\pT$ and $\nch>20$ region, the values of $|\rhon{2}|$ are $\lesssim 0.02$ in the PbPb and $\lesssim 0.06$ in the $p$Pb collisions, but are significantly larger in the $pp$ collisions ($\sim 0.1-0.2$). For $\rhon{3}$, current statistical uncertainties do not provide a precise lower limit for the non-flow contributions in $p$Pb and PbPb collisions. But in $pp$ collisions and higher $\pT$, the non-flow effects could lead to $\rhon{3}$ values significantly larger than one.

\begin{figure}[h!]
\begin{center}
\includegraphics[width=0.7\linewidth]{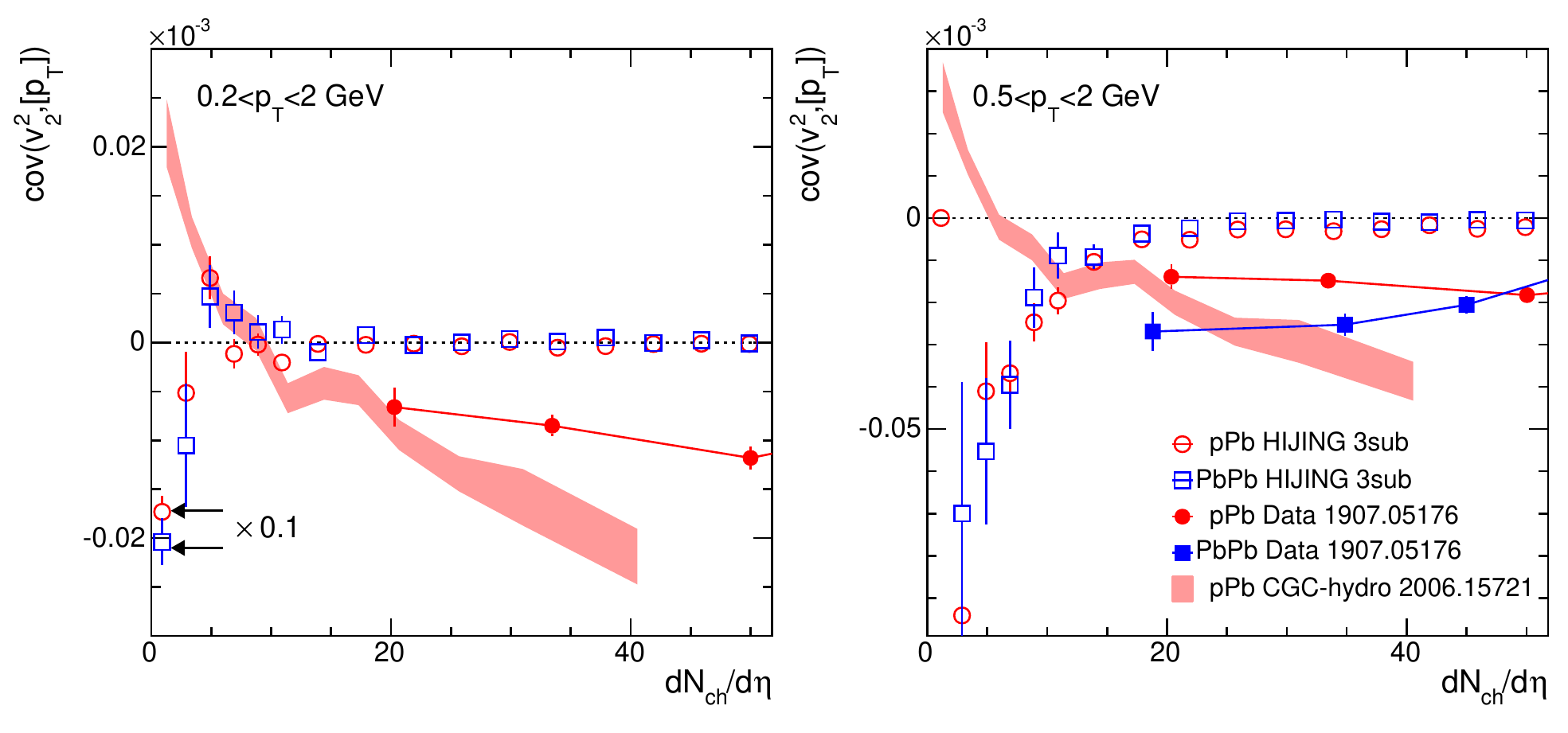}\\
\includegraphics[width=0.7\linewidth]{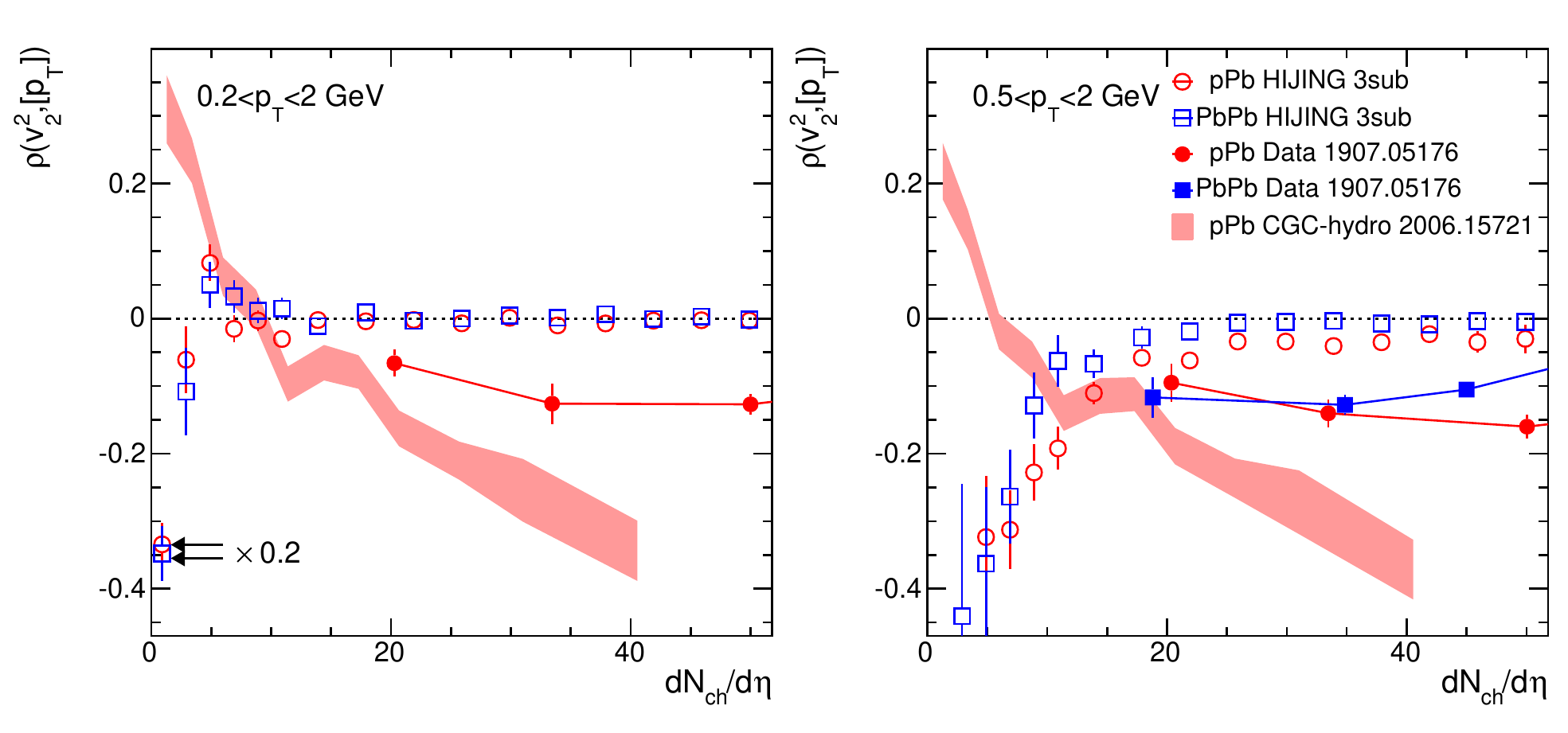}
\end{center}
\caption{\label{fig:6} The $\cov{2}$ (top) and $\rhon{2}$ (bottom) as a function of $\nch$  in $p$Pb and PbPb collisions for $0.2 < \pT < 2$ GeV (left) and $0.5 < \pT < 2$ GeV (right). The results are obtained from the three-subevent method and compared with experimental data from ATLAS and CGC-hydro model calculations~\cite{Giacalone:2020byk} that include the initial-state momentum correlations. The data points with $\nch<10$ in the left panels have been rescaled by the factors in order to fit into the y-ranges.}
\end{figure}

Equipped with these detailed knowledge of non-flow, we are ready to discuss its impact on the interpretation of the $v_n$--$[\pT]$ correlation in terms of ISM and FSM. The top panels of Figure~\ref{fig:6} compare the non-flow expectation of $\cov{2}$ with the ATLAS data~\cite{Aad:2019fgl}~\footnote{The $x$-axis of ATLAS data corresponds to number of charged particles in $|\eta|<2.5$ and 0.5-5 GeV, which needs to be multiplified by 2/5 to convert to the $\nch$ value. The factor of 5 corresponds to the rapidity range and the factor of 2 is covertion to multiplicity in $\pT>0.1$ GeV.}. The strength of the non-flow correlations is much smaller than the experimental data in the PbPb collisions (which covers $\nch>20$ region), but could be significant in $p$Pb collisions in $0.5<\pT<2$ GeV, reaching a level of around 30--40\% of the experimental values at $\nch\sim20$. The results are also compared to the CGC-hydro model for $p$Pb~\cite{Giacalone:2020byk} that includes both ISM and FSM but without non-flow. In the $\nch>20$ region where the FSM dominates, the model over-predicts the experimental data. In the $\nch<10$ region, the CGC-hydro  model is dominated by a positive ISM signal, which seems to be smaller in magnitude than the expected non-flow contribution. It might be that the combined non-flow and ISM would still remain negative for the $0.5<\pT<2$ GeV range. For the $0.2<\pT<2$ GeV range where the non-flow contribution is smaller, the combined signal could be slightly positive around $\nch\sim5-10$, but would still remain negative at $\nch\sim5$. 

The bottom panels of Figure~\ref{fig:6} show the same comparison in terms of $\rhon{2}$. The qualitative behaviors are largely the same, with a few important quantitative differences from $\cov{2}$. The non-flow contributions relative to the experimental data are larger, especially in the $p$Pb collisions, reaching more than 50\% of the experimental values at $\nch\sim20$ in $0.5<\pT<2$ GeV. This is due to the fact that the values of $\varp$ in HIJING are about a factor of 2 smaller than the experimental values, leading to a more negative $\rhon{2}$ closer to the data. We also caution that the ATLAS $\var{n}$ data, calculated via Eq.~\eqref{eq:6}, are still biased by non-flow contributions (see Appendix~\ref{sec:app1}), which reduce $\rhon{2}$ slightly further. The main message of Figure~\ref{fig:6} is that the interpretation of the $\cov{2}$ at low $\nch$ region is rather complicated. Firstly, the non-flow contributions from our model studies are negative and could account for some of the observed negative signal in the low $\nch$ range that are also associated with the FSM. Secondly, the negative non-flow contributions compete with the ISM and may eliminate the sign-change in the actual measurement. Thirdly, the fact that $\PYTHIA$ model shows a positive $\rhon{2}$ at $\nch<10$ (Figure~\ref{fig:5}) suggests that the sign of non-flow contributions is model-dependent and could also be positive. In the latter case, even if experiments observe a positive $\rhon{2}$, one could not easily interpret this signal as generated by the ISM.

\section{Summary}\label{summary}
The influences of non-flow effects to the three-particle correlation between harmonic flow $v_n$ and event-by-event average transverse momentum $[\pT]$, $\cov{n}$, are studied in $pp$, $p$Pb and peripheral PbPb collisions for $n=2-4$. This study is performed using $\PYTHIA$ and $\HIJING$ event generators, which contain only non-flow correlations such as fragmentation of jet and dijets and resonance decays, but have no genuine long-range multi-particle correlations from the initial-state or the final-state evolution. The efficacy of non-flow suppression via the rapidity separated three-subevent method has been tested, and is observed to give smallest $|\cov{n}|$ values in comparison to the standard and two-subevent methods for all harmonics, collision systems and $\pT$ ranges investigated in this paper. The values of $\cov{2}$ from the three-subevent method are negative in the region $\nch>20$ and approach zero towards higher $\nch$. The magnitudes of the $\cov{2}$ are much smaller than the experimentally measured values in the $p$Pb and PbPb collisions, suggesting that the measured $\cov{2}$ values in $\nch>20$ reflect genuine correlations arising from the final-state interactions. In the region $\nch<20$, the values of $\cov{2}$ decrease toward more negative values in $\HIJING$ simulations of $p$Pb and PbPb collisions, but increases in $\HIJING$ and $\PYTHIA$ simulations of $pp$ collisions. They reach a maximum (positive for $\PYTHIA$ but is negative in $\HIJING$) at around $\nch\sim20$ before decreasing again for $\nch<5$. The differences between $\HIJING$ and $\PYTHIA$ suggest that the non-flow contributions in $\nch<20$ region are highly model-dependent. The predicted sign change of $\nch$ from initial-state momentum correlation due to gluon saturation physics may not be observed if the non-flow contributions are negative, or unambiguous if the non-flow contributions are positive. Further detailed quantitative model investigation of these different sources are required.

We appreciate comments from  G. Giacalone, S. Huang, J. Nagle, and B. Schenke. We thank B. Schenke for sharing the CGC-hydro model calculation, and J. Nagle'suggestion to compare $pp$ collisions between $\PYTHIA$ and $\HIJING$. This work is supported by DEFG0287ER40331 and PHY-1913138.
%\clearpage
\appendix
\section{Influence of non-flow on $\var{n}$}\label{sec:app1}
\begin{figure}[h!]
\begin{center}
\includegraphics[width=0.9\linewidth]{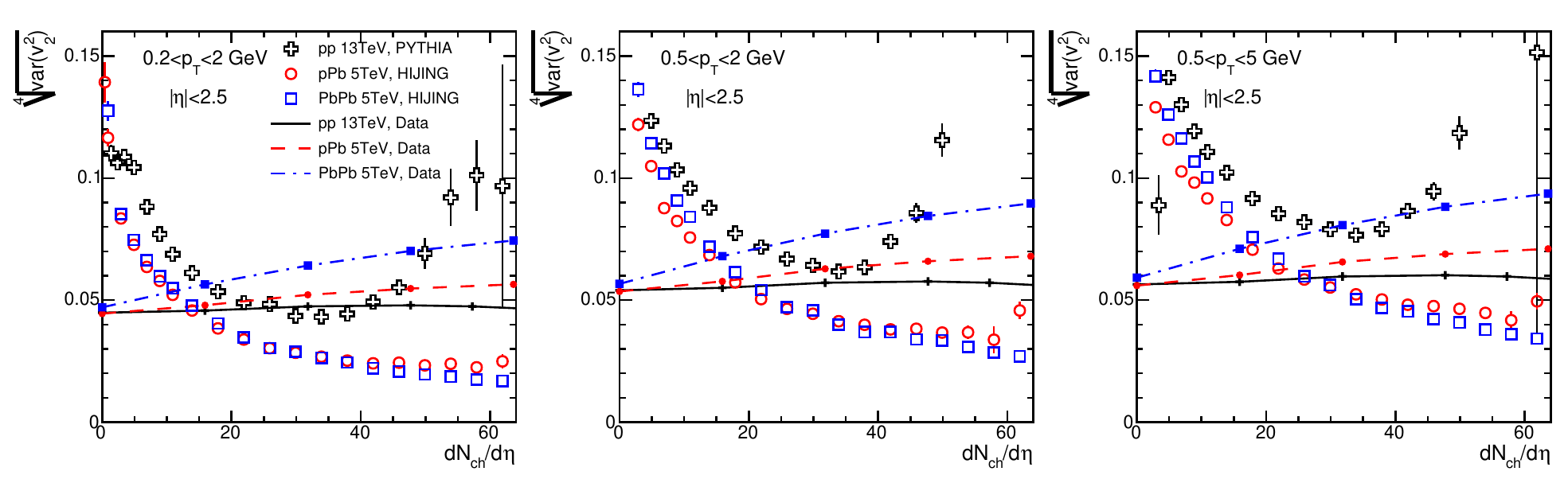}\\
\includegraphics[width=0.9\linewidth]{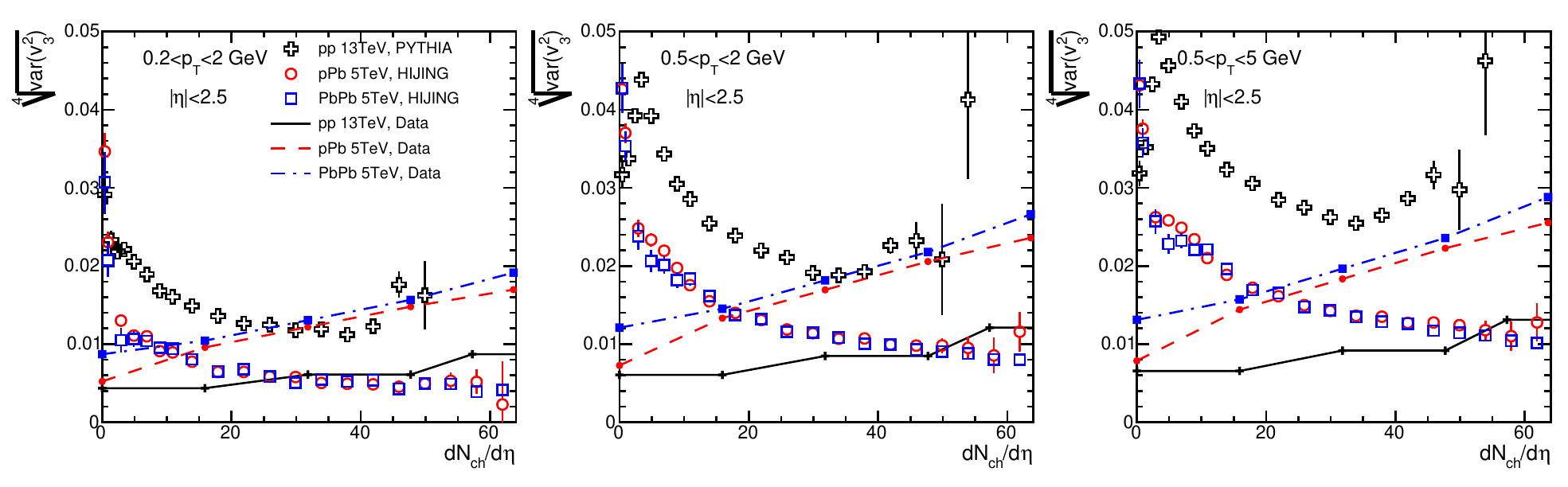}
\end{center}
\caption{\label{fig:app1} The $\sqrt[4]{\var{2}}$ (top) and $\sqrt[4]{\var{3}}$ (bottom) as a function of $\nch$ compared between three collision systems for $0.2 < \pT < 2$ GeV (left), $0.5 < \pT < 2$ GeV (middle), and $0.5 < \pT < 5$ GeV (right). The data points are calculated from $\PYTHIA$ and $\HIJING$ models via Eq.~\eqref{eq:6} and the lines are estimated via Eq.~\eqref{eq:7} from published $v_n$ data.}
\end{figure}

In the ATLAS measurement~\cite{ATLAS:2021kty}, the $\var{n}$ was calculated using Eq.~\eqref{eq:6}. In the low $\nch$ region, the $c_n\{2\}_{\mathrm{2sub}}$ and the resulting $\rhon{n}$ could be strongly biased by the non-flow correlations. Figure~\ref{fig:app1} compares $\var{n}$ from Eq.~\eqref{eq:6} with those estimated via Eq.~\eqref{eq:7} based on published $v_n$ data in three collision systems. They are presented in terms of $\sqrt[4]{\var{n}}$ in order to be shown in the familiar scale as the single-particle $v_n$ values. In $p$Pb and PbPb collisions, the non-flow is sub-dominant for $\nch>20$ but can be larger than the genuine flow signal at lower $\nch$ values. In the $pp$ collisions, the non-flow contribution is comparable or larger than the genuine flow signal over the full $\nch$ range. 

\begin{figure}[h!]
\begin{center}
\includegraphics[width=1.0\linewidth]{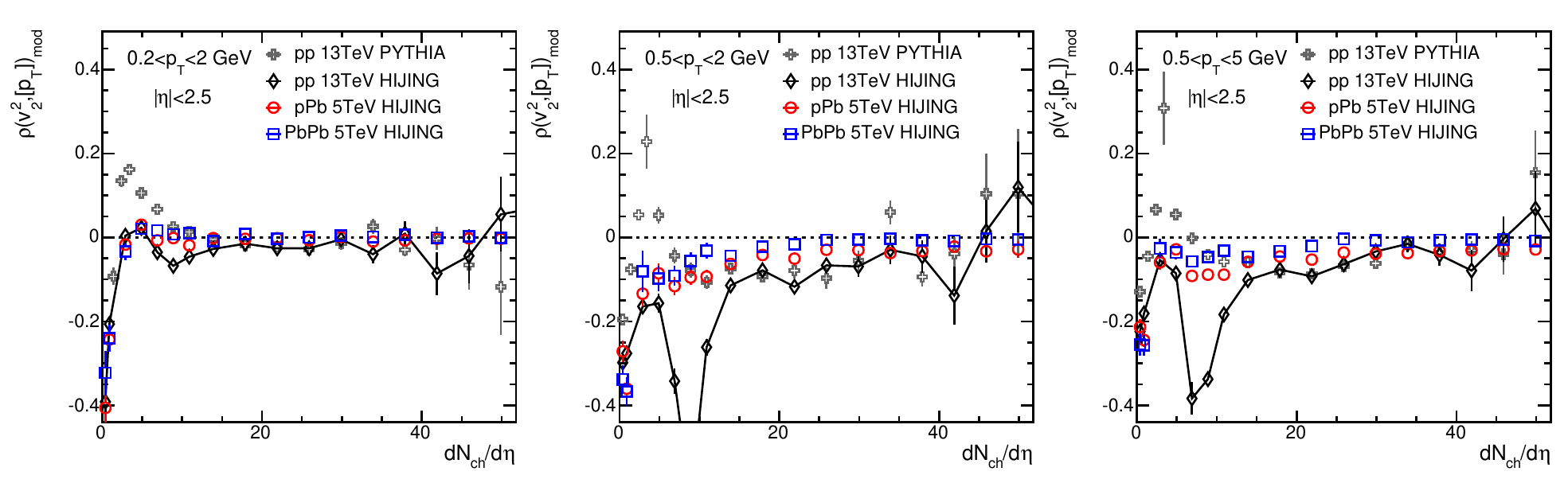}
\includegraphics[width=1.0\linewidth]{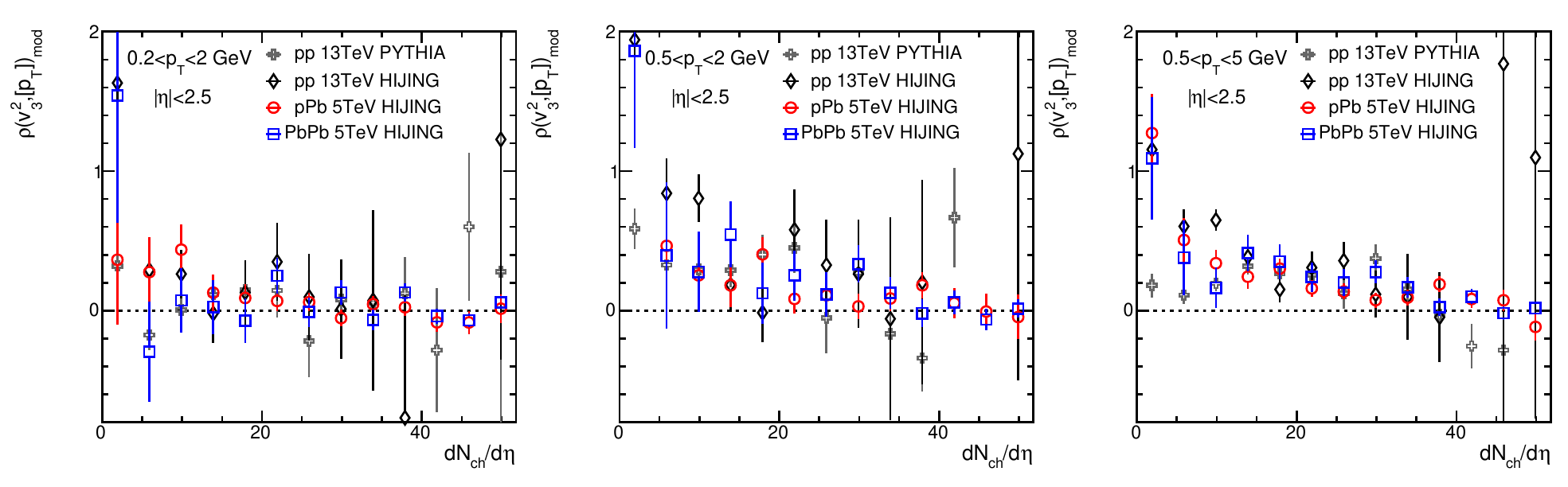}
\end{center}
\caption{\label{fig:app2} The $\rhon{2}_{\mathrm{mod}}$ (top) and $\rhon{3}_{\mathrm{mod}}$ (bottom) as a function of $\nch$ from the three-subevent method compared between three collision systems for $0.2 < \pT < 2$ GeV (left), $0.5 < \pT < 2$ GeV (middle), and $0.5 < \pT < 5$ GeV (right). The results are calculated using the modified form of $\var{2}$ that includes both the flow and non-flow via the procedure described in the text.}
\end{figure}

To estimate the possible bias of the non-flow, we add the $\var{n}$ from flow and non-flow of Figure~\ref{fig:app1} in quadrature sum: $\var{n}_{\mathrm{mod}}=\sqrt{\var{n}_{\mathrm{flow}}^2+ \var{n}_{\mathrm{non-flow}}^2}$. The $\var{n}_{\mathrm{mod}}$ is then used to obtain a modified form of Pearson coefficient $\rhon{n}_{\mathrm{mod}}$. The results are shown in Figure~\ref{fig:app2}. Comparing to the original unbiased results in Figure~\ref{fig:5}, the magnitudes of the $\rhon{n}_{\mathrm{mod}}$ are much reduced in the low $\nch$ region due to the large non-flow bias to $\var{n}$. The differences between the three systems are also artificially reduced. Therefore, it is important to use a $\var{n}$ that is free of non-flow effects by following the procedure given in Eq.~\eqref{eq:7}. 

\bibliography{nonflow}{}
\bibliographystyle{apsrev4-1}

\end{document}